\documentclass[sigconf]{acmart}

\usepackage{booktabs} 

\copyrightyear{2017}
\acmYear{2017}
\setcopyright{acmlicensed}
\acmConference{SC17}{November 12--17, 2017}{Denver, CO, USA}
\acmPrice{15.00}
\acmDOI{10.1145/3126908.3126927}
\acmISBN{978-1-4503-5114-0/17/11}

\usepackage[T1]{fontenc}
\usepackage{color}

\usepackage[binary-units=true]{siunitx}
\usepackage{algorithm2e}
\DeclareSIUnit\parsec{pc}
\begin{document}


\title{Galactos: Computing the Anisotropic 3-Point Correlation Function for 2 Billion Galaxies}

\author{Brian Friesen}
\affiliation{%
\institution{Lawrence Berkeley National Laboratory}
\streetaddress{1 Cyclotron Road, M/S 59R4010A}
\city{Berkeley}
\state{CA}
\postcode{94720}
}

\author{Md. Mostofa Ali Patwary}
\affiliation{ %
\institution{Intel Corporation}
\streetaddress{2200 Mission College Blvd.}
\city{Santa Clara}
\state{CA}
\postcode{95054}
}

\author{Brian Austin}
\affiliation{%
\institution{Lawrence Berkeley National Laboratory}
\streetaddress{1 Cyclotron Road, M/S 59R4010A}
\city{Berkeley}
\state{CA}
\postcode{94720}
}

\author{Nadathur Satish}
\affiliation{ %
\institution{Intel Corporation}
\streetaddress{2200 Mission College Blvd.}
\city{Santa Clara}
\state{CA}
\postcode{95054}
}

\author{Zachary Slepian}
\affiliation{%
\institution{Einstein Fellow, Lawrence Berkeley National Laboratory}
\streetaddress{1 Cyclotron Road, M/S 50R6048}
\city{Berkeley}
\state{CA}
\postcode{94720}
}

\author{Narayanan Sundaram}
\affiliation{ %
\institution{Intel Corporation}
\streetaddress{2200 Mission College Blvd.}
\city{Santa Clara}
\state{CA}
\postcode{95054}
}

\author{Deborah Bard}
\affiliation{%
\institution{Lawrence Berkeley National Laboratory}
\streetaddress{1 Cyclotron Road, M/S 59R4010A}
\city{Berkeley}
\state{CA}
\postcode{94720}
}

\author{Daniel J. Eisenstein}
\affiliation{%
\institution{Harvard-Smithsonian Center for Astrophysics}
\streetaddress{60 Garden Street, MS-20,}
\city{Cambridge}
\state{MA}
\postcode{02138}
}

\author{Jack Deslippe}
\affiliation{%
\institution{Lawrence Berkeley National Laboratory}
\streetaddress{1 Cyclotron Road, M/S 59R4010A}
\city{Berkeley}
\state{CA}
\postcode{94720}
}

\author{Pradeep Dubey}
\affiliation{ %
\institution{Intel Corporation}
\streetaddress{2200 Mission College Blvd.}
\city{Santa Clara}
\state{CA}
\postcode{95054}
}

\author{Prabhat}
\affiliation{%
\institution{Lawrence Berkeley National Laboratory}
\streetaddress{1 Cyclotron Road, M/S 59R4010A}
\city{Berkeley}
\state{CA}
\postcode{94720}
}

\renewcommand{\shortauthors}{B. Friesen et al.}

\begin{abstract}

    The nature of dark energy and the complete theory of gravity are two central questions currently facing cosmology. A vital tool for addressing them is the 3-point correlation function (3PCF), which probes deviations from a spatially random distribution of galaxies. However, the 3PCF's formidable computational expense has prevented its application to astronomical surveys comprising millions to billions of galaxies. We present Galactos, a high-performance implementation of a novel, $\mathcal{O}(N^2)$ algorithm that uses a load-balanced k-d tree and spherical harmonic expansions to compute the anisotropic 3PCF. Our implementation is optimized for the Intel Xeon Phi architecture, exploiting SIMD parallelism, instruction and thread concurrency, and significant L1 and L2 cache reuse, reaching 39\%\ of peak performance on a single node. Galactos scales to the full Cori system, achieving 9.8~PF (peak) and 5.06~PF (sustained) across 9636 nodes, making the 3PCF easily computable for all galaxies in the observable universe.

\end{abstract}

\maketitle

\section{Overview}
\label{sec:overview}
\subsection{Probing dark energy, gravity, and galaxy formation}
Measuring the anisotropic 3PCF can illuminate two of the most significant challenges in present-day  cosmology: dark energy's fundamental nature, and the complete theory of gravity. Dark energy drives accelerated expansion of the Universe. It constitutes 72\% of the Universe's current energy density, but its fundamental nature remains unknown. Unlike all other known substances, dark energy has negative pressure, and it is not predicted by the Standard Model of particle physics.

An alternative explanation for the accelerated expansion of the Universe is that the most widely accepted theory of gravity, General Relativity (GR), requires modification. Indeed, we already know that GR must be only an approximation to the true theory of gravity because GR cannot be unified with quantum field theory, which governs the interactions of the fundamental particles \cite{Copeland}.

The clustering of galaxies throughout space offers the largest possible laboratory  for pursuing these fundamental questions about the Universe's contents and governing laws. Galaxy clustering is often quantified via correlation functions. These measure the excess of pairs of galaxies as a function of the distance between the galaxies (2-point correlation function, 2PCF) or the excess of triplets as a function of triangle configuration (3PCF) compared to a random distribution \cite{Peebles}.

The 2PCF has been a highly successful tool in the past for exploring both dark energy and gravity.  In particular, 
the Baryon Acoustic Oscillation (BAO) method uses a sharp feature in the 2PCF  as a ``standard ruler'' \cite{Eisenstein1998}. Locating this feature in the 2PCF of galaxy samples from different epochs in the Universe's history enables the mapping of the Universe's expansion over time,  in turn illuminating the nature of dark energy \cite{Weinberg,Alam2016}.

Further, the growth rate of structure can be probed using the { \it{anisotropic}} (direction-dependent) 2PCF. This tracks the excess pairs of galaxies compared to a random distribution as a function of both the separation between the galaxies and the angle between the separation vector and the line of sight to the galaxy pair. 

While the underlying clustering of galaxies is independent of angle to the line of sight, the clustering {\it observed} in a redshift survey is modulated because of Redshift Space Distortions (RSD)~\cite{HamiltonRSD}. RSD occur because galaxies' own (``peculiar'') velocities with respect to the background expansion of the Universe affect our inference of their positions along the line of sight from their redshifts.  

Galaxies' velocities are generated by growth of structure due to gravity, and GR makes a particular prediction for the growth rate. Any measured deviation would offer a vital clue to the modifications required for a complete theory of gravity \cite{Lindergrowth}.

\subsection{Unlocking the anisotropic 3PCF's potential}
In principle, the 3PCF offers a new lever to understand both dark energy and gravity.  
\cite{Sefasutti2006} demonstrated that adding 3PCF information to an analysis of the 2PCF can improve constraints on the cosmological parameters that describe the nature of the universe. 
Measurements of the amount of dark energy growth and the rate of structure (which constrains gravity) can be improved by up to a factor of 2, and biasing, which describes how galaxies trace the underlying matter and is vital for understanding galaxy formation \cite{Fry1994,SERV,GilMarin2}, can be improved by a factor of $~$2 compared to using the 2PCF alone.
Like the 2PCF, the 3PCF has BAO features that can be used as a standard ruler to trace the expansion history \cite{SERSDmodel,SEBaoDetxn}.  And similarly to the anisotropic 2PCF, the anisotropic 3PCF contains valuable information on the growth rate. 

\begin{figure}
\centering
\includegraphics[width=\columnwidth]{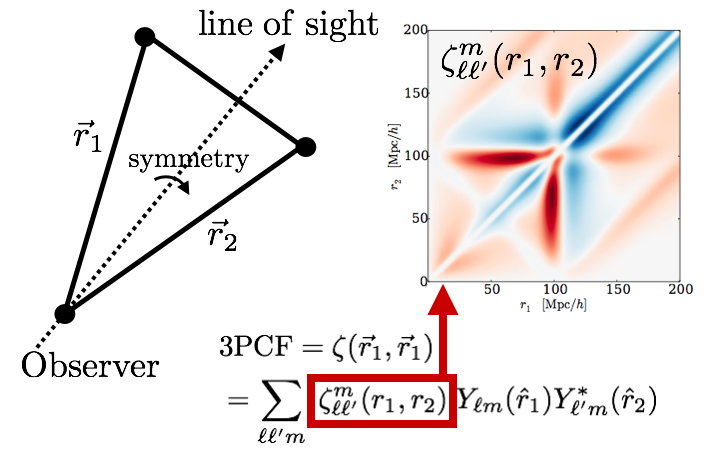}
\caption{Representation of a triangle configuration for the 3PCF. Each dot represents a galaxy. The anisotropic 3PCF depends on the vectors $\vec{r}_1$ and $\vec{r}_2$, the relative distances to the primary galaxy, which is at the bottom left vertex. The relevant quantities are the triangle side lengths $r_1$ and $r_2$, the angle between $\vec{r}_1$ and $\vec{r}_2$, and $\vec{r}_1$ and $\vec{r}_2$'s angles to the line of sight (dashed arrow). We expand the angular dependence of the anisotropic 3PCF in the basis of spherical harmonics, with the dependence on triangle side lengths $r_1$ and $r_2$ encoded in the radial coefficient $\zeta^m_{\ell \ell'}(r_1,r_2)$. The panel on the right (taken from~\cite{SE3ptalg}) shows a schematic of the algorithm's output: a coefficient $\zeta_{\ell \ell'}^m$ as a function of the triangle side lengths $r_1$ and $r_2$. The color indicates the number of triangles; red is an excess over a spatially random distribution and blue a deficit. The features are from BAO. 
\label{fig:alg_plus_eqn}}
\end{figure}

The anisotropic 3PCF depends on two triangle sides, their enclosed angle, and the angle of each side to the line of sight to the galaxy triplet (see Figure \ref{fig:alg_plus_eqn}). Consequently it has much richer structure than the anisotropic 2PCF, and offers many different configurations that all ultimately probe gravity \cite{RampfWong,Scoccimarro} as well as dark energy. 
It has never been measured. Not only would it provide additional information compared to the anisotropic 2PCF, but it would break significant degeneracies between the growth rate and other cosmological parameters that cannot be broken by the anisotropic 2PCF alone.

Finally, the 3PCF is a highly sensitive measure of how galaxies form. Galaxies are termed a ``biased'' tracer of the underlying dark matter density: they do not follow it with perfect fidelity, but rather form based on local conditions, such as the value of the local gravitational tidal forces or the relative velocity between regular matter and dark matter \cite{McDonald2009, TH2010}. 

 These effects enter at sub-leading order in the 2PCF, meaning one is searching for a small change to a large baseline signal. However, in the 3PCF, these effects are at leading order, making the 3PCF a uniquely powerful probe of galaxy biasing \cite{SERV,SERSDmodel}.

\subsection{The 3PCF is computationally demanding}
Unfortunately, in practice, the 3PCF has rarely been used to constrain models of cosmology: counting all possible triangles formed by galaxy triplets in a modern redshift survey is combinatorially explosive. It naively scales as $\mathcal{O}(N^3)$ compared to $\mathcal{O}(N^2)$ scaling for the 2PCF, where $N$ is the number of galaxies in the survey. The 3PCF and bispectrum (its Fourier-space analog) have had other uses, but given this combinatoric challenge have been restricted to only particular triangle configurations (e.g. isosceles) \cite{McBride, GilMarin1,GilMarin2}.  
Recent work with an algorithm \cite{SE3ptalg} related to the one used in this work has measured many triangle configurations for the spherically-averaged (isotropic) 3PCF \cite{SEBaoDetxn,SERV}, but the full anisotropic dependence of the 3PCF has never been measured. We note that the anisotropic 3PCF contains the isotropic 3PCF as a proper subset.

Until now no algorithm has existed to measure the full anisotropic 3PCF for all triangles, on large scales, and in million- to billion-galaxy surveys. In this work, we develop and implement an algorithm that measures the anisotropic 3PCF in $\mathcal{O}(N^2)$ time. Galactos, our high-performance parallel computing anisotropic 3PCF code will, for the first time, make the measurement of the 3PCF feasible for modern astronomical surveys. The mathematical framework of the algorithm and other details relevant for cosmology are presented more fully in a companion paper ~\cite{SE3ptAniso}.
Galactos' single node performance has been highly optimized for Intel Xeon Phi, achieving 39\%\ of peak single-node performance with efficient use of vectorization and the full memory hierarchy. 
Galactos presents almost perfect weak- and strong-scaling, and achieves 5.06~PF across 9636 nodes.

The only required input is the 3-D positions of the galaxies, which is already demanded by the 2PCF.  Thus, for zero additional cost in telescope time, our algorithmic and computing advances will yield significant additional insight on the most fundamental questions facing cosmology.

\section{Current State of the Art}

Up until quite recently, all existing 3PCF algorithms scaled as $\mathcal{O}(N^3)$. However, very recent work by~\cite{SE3ptalg} (summarized below) developed a new approach to the isotropic 3PCF, using Legendre polynomials and radial coefficients as a basis for the isotropic 3PCF's dependence on triangle opening angle and side lengths. This algorithm showed that using spherical harmonic (SH) expansions of the density field around each galaxy in the survey, the 3PCF's multipole moments (another name for the radial expansion coefficients) could be obtained scaling as $\mathcal{O}(N^2$).

In this section we first summarize developments previous to the Legendre approach. We then briefly discuss the Legendre approach for the isotropic 3PCF.
Finally, we outline previous work on high-performance implementations of the 2PCF and 3PCF.

\subsection{State of the art prior to Legendre polynomial approach}
Prior to the development of the Legendre polynomial approach, 3PCF algorithms \cite{Moore2001, Gray2004, Jarvis2004, Nichol2006, Gardner2007, March2013} fundamentally scaled as the number of possible triangles in a survey. More precisely, for a 3PCF measurement out to some maximum scale $R_{\rm max}$, there would be $N(nV_{R_{\rm max}})^2$ , with $n$ the survey number density and $V_{R_{\rm max}}$ the spherical volume within $R_{\rm max}$. 

\cite{Moore2001, Gray2004, Jarvis2004, Nichol2006, Gardner2007, March2013} trace the development of a 3PCF algorithm based on k-d trees, most recently described in \cite{March2013}. This algorithm used ``marked'' k-d trees that cached additional information: the number of galaxies within each node of the tree as well as the bounding box of the node. 

While somewhat faster than a naive triplet count, the k-d tree-based algorithms are most effective if the typical galaxy separation is much smaller than the desired binning in triangle side length used to report the 3PCF. If this is the case, then many galaxies can be handled at once using a node of the k-d tree without going further down the tree. 

However, typical galaxy surveys geared for cosmology are sparse. For example,  the Baryon Oscillation Spectroscopic Survey (BOSS) has an average separation between galaxies of $13~\text{Mpc}/h$, but the desired bin width is typically $\sim\! 10~\text{Mpc}/h$. Consequently k-d tree algorithms do not perform well for cosmological-scale 3PCF measurements. 
The largest-scale use-case tested in \cite{March2013} is a triangle configuration  with three sides of $5.6~\text{Mpc}/h$  each, significantly smaller than the large triangles ($\sim\!50-200~\text{Mpc}/h$) most useful for cosmology. Furthermore, these k-d tree algorithms still fundamentally scale as $N(nV_{\rm R_{\rm max}})^2\propto N^3$ \cite{March2013}.

\subsection{Legendre polynomial approach}

The current state of the art algorithm to measure the isotropic 3PCF is presented in \cite{SE3ptalg}. This algorithm formulates the isotropic 3PCF as
\begin{equation*}
\zeta(r_1, r_2;\hat{r}_1\cdot\hat{r}_2) = \sum_{\ell}\zeta_{\ell}(r_1,r_2)P_{\ell}(\hat{r}_1\cdot\hat{r}_2),
\end{equation*}
where $P_{\ell}$ is a Legendre polynomial. We wish to obtain the radial coefficients (``multipole moments'') $\zeta_{\ell}$, which describe the 3PCF's dependence on triangle side lengths $r_1$ and $r_2$.

The algorithm estimates the multipole moments around each galaxy in the survey, and averages them. For each galaxy in the survey, the algorithm bins the density around it into spherical shells, corresponding to bins in triangle side length. It then expands the angular clustering on each shell into spherical harmonics. Using the spherical harmonic addition theorem, one can then sum the spherical harmonics  over spins to recover the multipole moments. This algorithm scales as $\mathcal{O}(N^2)$.

We emphasize that the Legendre basis is symmetric under rotations by construction because they preserve the dot product $\hat{r}_1\cdot\hat{r}_2$. Thus the Legendre polynomial-based algorithm does not track {\it anisotropies} in the 3PCF. 

However, as  discussed in Section~\ref{sec:overview}, redshift-space distortions induce anisotropies that carry significant information on the growth rate of structure and hence the theory of gravity. The new algorithm implemented in this work  generalizes the Legendre polynomial approach to include anisotropies, as described in ~\cite{SE3ptAniso} and in Section~\ref{sec:a3PCF}.

\subsection{Current SoA implementations}
To date, the Legendre polynomial algorithm for the isotropic 3PCF provides the best baseline comparison for the current work. However we caution that the algorithm in this work tracks different quantities that offer additional information over the isotropic 3PCF, so the quantitative comparison should serve only as a guide.

The performance numbers for the Legendre algorithm quoted in \cite{SE3ptalg} used a dataset consisting of 642,619 randomly distributed particles in a realistic sky survey geometry; the calculation for the true distribution of galaxies would likely have similar performance. In that work, the isotropic 3PCF 
was run in 170 seconds on a 6-core 4.2 GHz i7-3930K. The core computation of the multipoles sustains about $30\%$ of peak performance for this processor, and the full code sustains about $21\%$ of peak. 
The implementation presented in \cite{SE3ptalg} included a number of optimizations to exploit AVX registers, as well as a simple gridding scheme to accelerate the finding of all secondary galaxies within $R_{\rm max}$ of a given primary.

In this work, we use our implementation of the novel anisotropic 3PCF algorithm  to process a dataset 3 orders of magnitude larger, on 4 orders of magnitude more nodes. There have been no published attempts to measure the 3PCF at scale on massively parallel architectures. 
However, we do note that previous work has exploited of GPU capabilities for the 2PCF, focusing on the brute force calculation of the distances between all galaxy pairs ~\cite{Bard2012,Ponce2012}. 
Unpublished work using GPUs to accelerate the brute-force 3PCF on GPUs also exists~\cite{Bard2014,Bellis2015}, but has not been scaled to datasets beyond 50,000 galaxies or beyond a single GPU. 

We are aware of one significant effort to optimize the 2PCF for HPC machines~\cite{Chhugani2012} that focused on a novel SIMD-friendly algorithm for histogram updates. This effort calculated the 2PCF for a dataset of 1.7 billion galaxies in 5.3 hours, using 25,600 Intel Xeon cores.

\section{Innovations}
We present Galactos, a scalable algorithm and highly optimized implementation for both the isotropic and anisotropic 3PCF. In this section we describe the innovations introduced in this implementation. 
These include a distinctive algorithm that reduces the computational complexity of the dominant kernel from  $\mathcal{O}(N^3)$ to $\mathcal{O}(N^2)$, code optimizations that extract maximum benefit from thread-level parallelism and vectorization to obtain $\sim\!39\%$ of the peak performance per node,
and an enhanced  distributed k-d tree partioning that scales to the full system size and provides good load balance. This code has been scaled to nearly 2 billion galaxies on 9636 nodes of the Intel Xeon Phi-based supercomputer ``Cori'' (Section \ref{sec:weak-scaling}).  We discuss the algorithmic and code optimizations that enabled this performance in this section.

\subsection{$\mathcal{O}(N^2)$ algorithm for the anisotropic 3PCF}
\label{sec:a3PCF}

Galactos uses a novel algorithm, described in detail in a companion paper \cite{SE3ptAniso}, that we implement here at massive scale. The algorithm builds on developments in~\cite{SE3ptalg}  but generalizes to include fundamentally new, additional information.

The anisotropic 3PCF depends on two triangle sides, $r_1$ and $r_2$, the angle between them, 
and the angle of each triangle side to the line of sight to the galaxy, which we here take to be the $z$-axis. 
Note that these triangle sides define the radial bin, as illustrated in  Figure~\ref{fig:alg_plus_eqn}. 

The unit vectors describing the sides' directions will capture both the angle between the two sides and the angle between each side and the line of sight.  Spherical harmonics provide a complete basis for directions, so we may write the 3PCF as
\begin{equation*}
\zeta(\vec{r}_1,\vec{r}_2)=\sum_{\ell \ell' m}\zeta_{\ell\ell'}^m(r_1,r_2)Y_{\ell m}(\hat{r}_1)Y^*_{\ell' m}(\hat{r}_2).
\end{equation*}
Note that the $m$ (spin) of the two spherical harmonics are equal---this results from the axisymmetry about the $z$-axis (line of sight). 

The algorithm proceeds by measuring the 3PCF for all galaxies in the survey out to a maximum scale $R_{\text{max}}$, iterating over each "primary" galaxy; the first step for a given primary galaxy is to obtain all neighbors (secondaries) within that sphere. 
We use $R_{\text{max}} = 200~\text{Mpc}/h$ ($1~\text{Mpc}/h \approx 4.59$ million light-years). On scales larger than $200~\text{Mpc}/h$  there are too few independent samples of galaxies available in the Universe to add meaningful information (e.g. \cite{SERSDmodel}).

The secondaries are then binned into spherical shells based on distance from the primary; this corresponds to the bins in triangle side lengths $r_1$ and $r_2$. 

\begin{figure}
\centering
\includegraphics[width=\columnwidth]{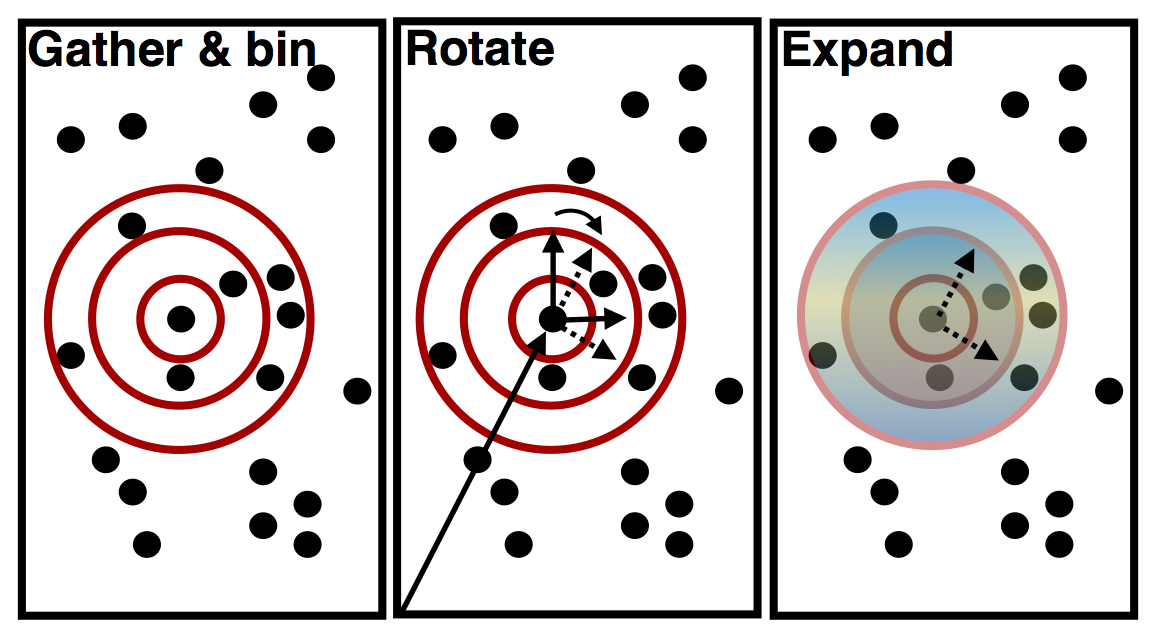}
\caption{About a given primary galaxy, the algorithm first gathers all neighbors (``secondaries'') within $R_{\rm max}$ and bins them into spherical shells. Second, it rotates all coordinates so that the primary is along the $z$-axis of the line of sight to the observer, and all of the secondaries' separation vectors from the primary are transformed to that frame. Third, the algorithm expands the angular dependence of the galaxies within each radial bin into spherical harmonics; this expansion is represented here by the shading. 
\label{fig:alg_flow}}
\end{figure}

In order to track the anisotropic clustering, the key step is to rotate the primary and all secondaries associated with that primary such that the primary lies on the $z$-axis of the line of sight. 
This process is illustrated in Figure~\ref{fig:alg_flow}. 
We also provide pseudo-code describing the implementation of this algorithm in Algorithm~\ref{algo:3pcf}.

We then need to obtain the spherical harmonic coefficients $a_{\ell m}$ of the secondaries' angular clustering on each spherical shell.  The desired radial coefficient $\zeta_{\ell \ell'}^m$ estimated about the given primary is simply $a_{\ell m}(r_1)a_{\ell' m}^*(r_2)$ for a bin combination of triangle side lengths $r_1,r_2$. 

To obtain these coefficients, we are calculating the integral
\begin{equation*}
\int d\Omega Y_{\ell m}(\hat{r})\delta(\hat{r}),
\end{equation*}
where $\Omega$ is the solid angle and $\delta$ is the density as a function of angle of a given bin. This integral reduces to a discrete sum for a set of points, becoming $\sum_i Y_{\ell m}(\hat{r}_i)$ where $i$ indexes all galaxies in a given radial bin.  Thus we simply evaluate the spherical harmonics at the location of each galaxy in the bin.

To do so, we construct the scaled separations $\Delta x/r,\Delta y/r,$ and $\Delta z/r$ between each secondary and the primary, where
\begin{equation*}
r=\sqrt{\Delta x^2 +\Delta y^2 +\Delta z^2}
\end{equation*}
is the distance between the secondary and the primary. The spherical harmonic coefficients are then just weighted sums of all combinations of powers of these
\begin{equation}
\left( \frac{\Delta x}{r} \right)^k \left( \frac{\Delta y}{r} \right)^p \left(\frac{\Delta z}{r} \right)^q
\label{eqn:powers}
\end{equation}
such that $k+p+q<=\ell$.  
This procedure yields the estimated anisotropic 3PCF about a single primary; it is repeated for all primaries.  

\begin{algorithm}
    \SetAlgoLined
    \KwData{$(x,y,z)$ coordinates of a set of galaxies}
    \KwResult{Estimation of 3-point correlation function}
    \For{$i \leftarrow 1$ to $N_{\text{primaries}}$} {
        \If {not primary} {skip galaxy}
        Search node-local k-d tree for neighbors (secondaries) of $i$ within $r_{\text{max}}$\;
        \For{$j \leftarrow 1$ to $N_{\text{secondaries}}$} {
            compute Cartesian distances $(\Delta x, \Delta y, \Delta z)$ between $i$ and $j$\;
            rotate distances\;
            compute radial distance $r$ between $i$ and $j$\;
            compute radial bin $r_n$ for galaxy pair $(i, j)$\;
            compute power combinations $((\Delta x/r)^a, (\Delta y/r)^b, (\Delta z/r)^c)$\;
            add power combinations to multipole for $r_n$\;
        }
    } 
    
    \caption{Pseudo-code describing the main computational kernel of Galactos.}
    \label{algo:3pcf}
\end{algorithm}

\subsection{Multi-node scaling}
\label{subsec:multinode_scaling}

The distribution of matter in the observable universe exhibits a complex variety of structures, on multiple length scales up to approximately \num{200}~\si{\mega\parsec}$/h$.
Partitioning such a problem across distributed memory computers requires a non-uniform partition of data in order to avoid significant load imbalances.
Our initial implementation used the parallel k-d tree partitioning from \cite{patwary2015bd}, which recursively partitions MPI processes and spatial domains into two sub-communicators with equal numbers of galaxies (but unequal volumes) until there is one sub-space per process.
The requirement of a perfect binary tree constrained the concurrency of the original algorithm to powers of two.

To overcome this restriction, we modified the k-d tree partitioning scheme so that each level of the tree divides MPI processes into sub-communicators of nearly equal size (i.e., equal to within a factor of 2) and divides galaxies in proportion to the sizes of the sub-communicators.
This important change enabled us to relax the constraint of using power-of-two nodes that exists in~\cite{patwary2015bd}. 
In our case, we were able to use 9636 nodes on Cori, rather than being restricted to 8192 nodes.

Computation of the 3PCF requires evaluation of the relative coordinates between each primary galaxy and all other galaxies within a cutoff distance.
We avoid inter-process communication \emph{ during} the 3PCF evaluation by exchanging all necessary neighbor galaxies beforehand.
This step is analogous to the ``halo exchange'' encountered in many structured grid PDE solvers, but is complicated by the irregular partitioning, which prevents \textit{a priori} computation of a process's neighbor list.
We perform the halo exchange by following the k-d partition: for each branch of the tree, a process gathers galaxies within the cutoff radius from the partition boundary, and sends copies of these particles to a peer on the opposite sub-communicator. 
Note that we split our dataset evenly across all available nodes on Cori (see Section~\ref{sec:outer-rim} for details), so each node processes 225,000 primaries. This corresponds to a box roughly $140~\text{Mpc}/h$ on a side, so to contain all secondaries that lie within $200~\text{Mpc}/h$ almost 2 million additional galaxies must be assigned to each node during the halo exchange.  
The remainder of the 3PCF calculation (besides a final reduction) is strongly parallel, leading to near-perfect weak scaling, as we show in Section~\ref{sec:weak-scaling}.

The overall load balance is determined by the number of pairs of primary and secondary (halo) galaxies on each node.
The number of primary galaxies is explicitly balanced by the k-d tree partition.
Because our cutoff distance for our 3PCF is large (\num{200}~\si{\mega\parsec}$/h$), we can rely on the long-range homogeneity of density to balance the number of secondary galaxies per primary.
In practice, we observed load imbalances of $\sim\!25$\% in our weak scaling tests.

\subsection{Single-node optimizations}  
\label{subsec:single-node-optimizations}

The majority of the execution time in Galactos is spent in a doubly nested loop over primary and secondary galaxies in the dataset, as shown in Algorithm~\ref{algo:3pcf}.
Consequently, optimizations targeting this kernel affect code performance more than any other.
At the thread level, we achieve efficient parallelism by distributing the primary galaxies over threads. We use OpenMP dynamic scheduling to allocate primaries to threads, since we iterate over all the galaxies in the k-d tree and perform further computations only on primaries (ignoring secondary galaxies that are in the k-d tree because of halo exchange). Using a dynamic schedule gives a significant performance boost over using a static schedule. Each thread computes the distances to the secondaries and the multipole contributions independently; the multipole values are combined at the end of the loop over primary galaxies. This approach ensures maximum independent work for each thread. 
We now enumerate the finer-grained optimizations and parallelization strategies in the code.

\subsubsection{Pre-binning/post-binning}
\label{sec:binning}
The multipole algorithm accumulates combinations of powers of the distances between each pair of galaxies independently for each radial bin.
As a result, since the galaxy dataset is not sorted with respect to galaxy pair distances, consecutive pairs of galaxies in the innermost loop (the loop over secondary galaxies, see Algorithm~\ref{algo:3pcf}) contribute their multipole values to random radial bins, leading to inefficient memory access patterns.
Galactos mitigates this problem by collecting all pairs of one primary (i.e. pairs with all its secondaries) that fall in the same radial bin into temporary ``buckets'' of any desired size (to be set to fully exploit a given machine's vector registers).
When a bucket fills, then Galactos computes the multipole contributions of all galaxies in that bucket.
This approach enables the use of effective vectorization over galaxy pairs, and also yields efficient cache reuse, since all vector operations access the multipole arrays corresponding to the same radial bin.
At the end of the loop over secondary galaxies, the buckets are swept once more, as they likely are only partially filled.

\subsubsection{Vectorization}
\label{subsubsec:vectorization}

Within the inner loop over secondaries (see Algorithm~\ref{algo:3pcf}), most of the code execution time is spent computing the multipole contributions of each pair (see  Figure~\ref{fig:single-node}).
Using a maximum multipole order $\ell = 10$ amounts to
\begin{equation*}
\frac{(\ell + 1) (\ell + 2) (\ell + 3)}{6} = 286
\end{equation*}
unique contributions of each galaxy pair to each radial bin's spherical harmonic expansion. 
One option is to vectorize this array of 286 elements; however

it is difficult to vectorize the computation along the multipole ``axis;'' doing so would lead to poor cache reuse due to the large number of temporary arrays that must be stored for each of the 286 power combinations.
However, without using these extra temporary arrays, the powers must be updated \textit{in situ}, leading to a data dependence which prohibits vectorization, e.g., the spherical harmonic multipole coefficient that is a combination $(\Delta x/r)^4(\Delta y/r)^5(\Delta z/r)^2$ is not computed until the value of $(\Delta x/r)^3(\Delta y/r)^5(\Delta z/r)^2$ is known.
Consequently, we vectorize the multipole computation loop over pairs rather than multipoles.

Within the multipole accumulation function for a given galaxy pair bucket for a given radial bin (as described in Section~\ref{sec:binning}), we compute the distance power combination given in Equation~\ref{eqn:powers}

from all galaxies to each multipole term $Y_{\ell m}$.
Ultimately these values are all accumulated to the same multipole array element, such that each group of 8 galaxy pairs (since 8 double precision numbers fill up the 512 bit-wide vector lanes on Xeon Phi) must perform a vector reduction of their contributions onto the same variable.
Thus for $N$ total galaxy pairs per radial bin, the algorithm must perform $N/8$ vector reductions per power combination.
To avoid computing so many reductions, we introduce an extra array of length 8 corresponding to each multipole $Y_{\ell m}$ for each radial bin.
Each time a set of 8 galaxy pairs computes their multipole contributions, they accumulate the result into the corresponding array locations of this temporary 8-element array.
After all galaxy pairs for a given primary galaxy have been computed, the code performs a single reduction of each of these 8-element arrays onto the corresponding multipole, replacing $N/8$ vector reductions with only 1 vector reduction for each of the 286 elements.

Computing the 286 multipole values given a single vector of $\Delta x/r, \Delta y/r, \Delta z/r$ involves many data dependencies; thus the amount of instruction-level parallelism (ILP) in this algorithm is limited.
In order to increase ILP, we perform computations on 4 independent vectors at once. 
This provides sufficient ILP to keep the hardware vector units busy. 
Note that this strategy increases register pressure and decreases performance if the number of independent vectors is increased beyond 4. 

Let us calculate the efficiency bound for the multipole addition function here.
For a set of $k$ pairs, we read in $3k$ elements ($\Delta x/r, \Delta y/r, \Delta z/r$), perform $286 \times 2k$ floating point operations and then load+store $286$ outputs. 
The flop to byte ratio is
\begin{equation*}
\frac{286\times 2 \times k}{(3k+286\times2)\times 8}
\end{equation*}
which is approximately equal to $1/8=0.125$ for small $k$ and $286\times 2/(3\times 8)=23.8$ for large $k$. 
We use a galaxy pair ``bucket'' size of $k=128$, giving a best-case flop/byte ratio of $9.6$. In practice, this ratio is slightly lower because of the need to load/store some temporary buffers due to blocking.

\subsubsection{Data Locality}

In the vectorized multipole computation loop, the distances for each galaxy pair are stored such that the $\Delta x$ for all pairs are stored contiguously, and similarly for the $\Delta y$ and $\Delta z$.

By storing the distances along a particular axis for all pairs contiguously, these vector operations result in the fewest possible number of loads from memory for the inputs.
Cache blocking is essential to achieving good performance on this kernel as the entire working set ($=128\times 8\times 3$ bytes (inputs) + $286\times 8\times 8$ (outputs) $= 21.4$ kB per thread) does not fit in L1 cache when run with 4 threads per core (see Section~\ref{sec:cori} for details of the memory configuration on Cori). Note that the outputs are each produced in an array of length 8 before a final reduction.
We compute a subset of the multipole terms for the entire bucket before moving on the next set of terms. This ensures that not all of the outputs need to be in cache at any given time, leading to better data locality.

\section{Problem Configuration}

In this section we describe the dataset configurations that were used to perform the weak- and strong-scaling measurements, as well as the Cori system itself on which these computations were performed.

\subsection{Description of Cori system}
\label{sec:cori}

The Galactos code was run on the Cori system at the National Energy Research Scientific Computing Center (NERSC) at Lawrence Berkeley National Laboratory.
Cori is a Cray XC40 system featuring 2,388 nodes of Intel Xeon Processor E5-2698 v3 (``Haswell'') and 9,688 nodes (recently expanded from 9,304) of Intel Xeon Phi Processor 7250 (``Knights Landing'').
All computations presented here were performed on Xeon Phi nodes.
Each of these nodes contains \num{68} cores (each supporting \num{4} simultaneous hardware threads), \num{16}~\si{\giga\byte} of on-package, multi-channel DRAM (``MCDRAM''), and \num{96}~\si{\giga\byte} of DDR4-2400 DRAM.
Cores are connected in a 2D mesh network with 2 cores per tile, and 1 MB cache-coherent L2 cache per tile. 
Each core has 32 KB instruction and 32 KB data in L1 cache. 
The nodes are connected via the Cray Aries interconnect.
In all Galactos computations presented here, we compiled the code using the Intel C++ v17.0.1 compiler with Cray MPI, and ran the code with 1 MPI process per Xeon Phi compute node, using 272 threads per node (4 threads per physical core).
Because the arithmetic intensity of our application is high, its memory bandwidth needs are modest; our measurements show that running the code with the Xeon Phi's high-bandwidth MCDRAM configured in ``flat'' mode (exposed as a separate NUMA domain from the DRAM) yields nearly identical performance as when the MCDRAM is configured in ``cache mode'' (configured as a transparent, direct-mapped cache to the DRAM).
As a result, all measurements reported in this work were performed with the MCDRAM in ``cache'' mode.

\subsection{Simulation data}
\label{sec:outer-rim}
\begin{figure}
\centering
\includegraphics[width=\columnwidth]{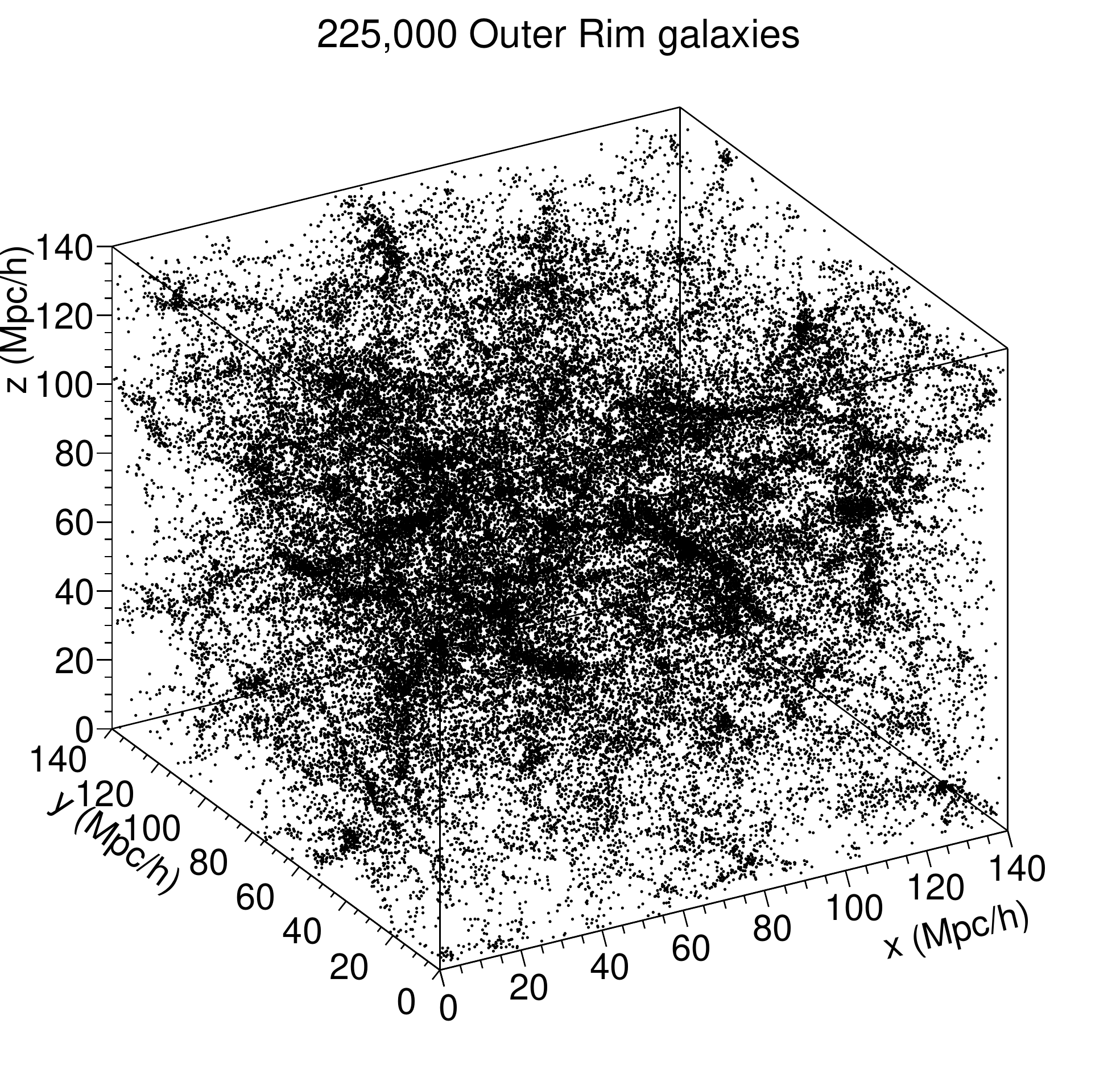}
\caption{Visualization of a box containing 225,000 galaxies in the Outer Rim simulation. 
\label{fig:outer-rim}}
\end{figure}

We ran the Galactos code over one of the largest cosmological simulation datasets available---the Outer Rim simulation \cite{Habib2016}. 
This simulation used over a trillion particles of dark matter, each of roughly $1.85~h^{-1}M_{\text{sun}}$, contained in a box of $3000~\text{Mpc}/h$ on each side. 
This distance corresponds to roughly 9.8 billion light-years, or $1/5$ the distance to the edge of the observable universe. 
The simulation was evolved over time, allowing structures to form in the distribution of the dark matter particles through gravitational attraction, and expansion to occur according to GR and dark energy with cosmological parameters close to the current concordance values \cite{WMAP7}.
We used a snapshot of the simulation at redshift $z=0$, i.e. the present day. The dark matter particles were grouped into gravitationally-bound "halos", which we take to represent galaxies (though the larger halos may actually host more than one galaxy). 
We used these $\sim\!2$ billion galaxies for our 3PCF analysis. 
If we partition these galaxies evenly between compute nodes on Cori (making a conservative estimate of 9000 nodes available) then each compute node is assigned 225,000 galaxies, visualized in Figure~\ref{fig:outer-rim}. 
To make partitions for our weak scaling tests, we selected cubes within Outer Rim  such that each cube enclosed the appropriate number of galaxies scaled down from the full 2 billion, as shown in Table~\ref{tab:weak_scaling_problem_sets}.

\begin{table}
\centering
\begin{tabular}{|c|c|c|}
\hline
    \# of nodes & \# of galaxies & cubic box length ($\text{Mpc}/h$)\\
\hline
\num{128} & \num{2.880e7} & \num{734.5}\\
\hline
\num{256} & \num{5.760e7} & \num{925.8}\\
\hline
\num{512} & \num{1.152e8} & \num{1166.9}\\
\hline
\num{1024} & \num{2.304e8} & \num{1470.9}\\
\hline
\num{2048} & \num{4.608e8} & \num{1853.3}\\
\hline
\num{4096} & \num{9.216e8} & \num{2334.7}\\
\hline
\num{8192} & \num{1.843e9} & \num{2934.4}\\
\hline
\num{9636} & \num{1.951e9} & \num{3000.0}\\
\hline
\end{tabular}
\caption{Datasets for weak scaling tests and full system run.}

\label{tab:weak_scaling_problem_sets}
\end{table}

\section{Performance Results}
\subsection{Single node performance}
\label{sec:single-node-performance}

\begin{figure}
\centering
\includegraphics[width=\columnwidth]{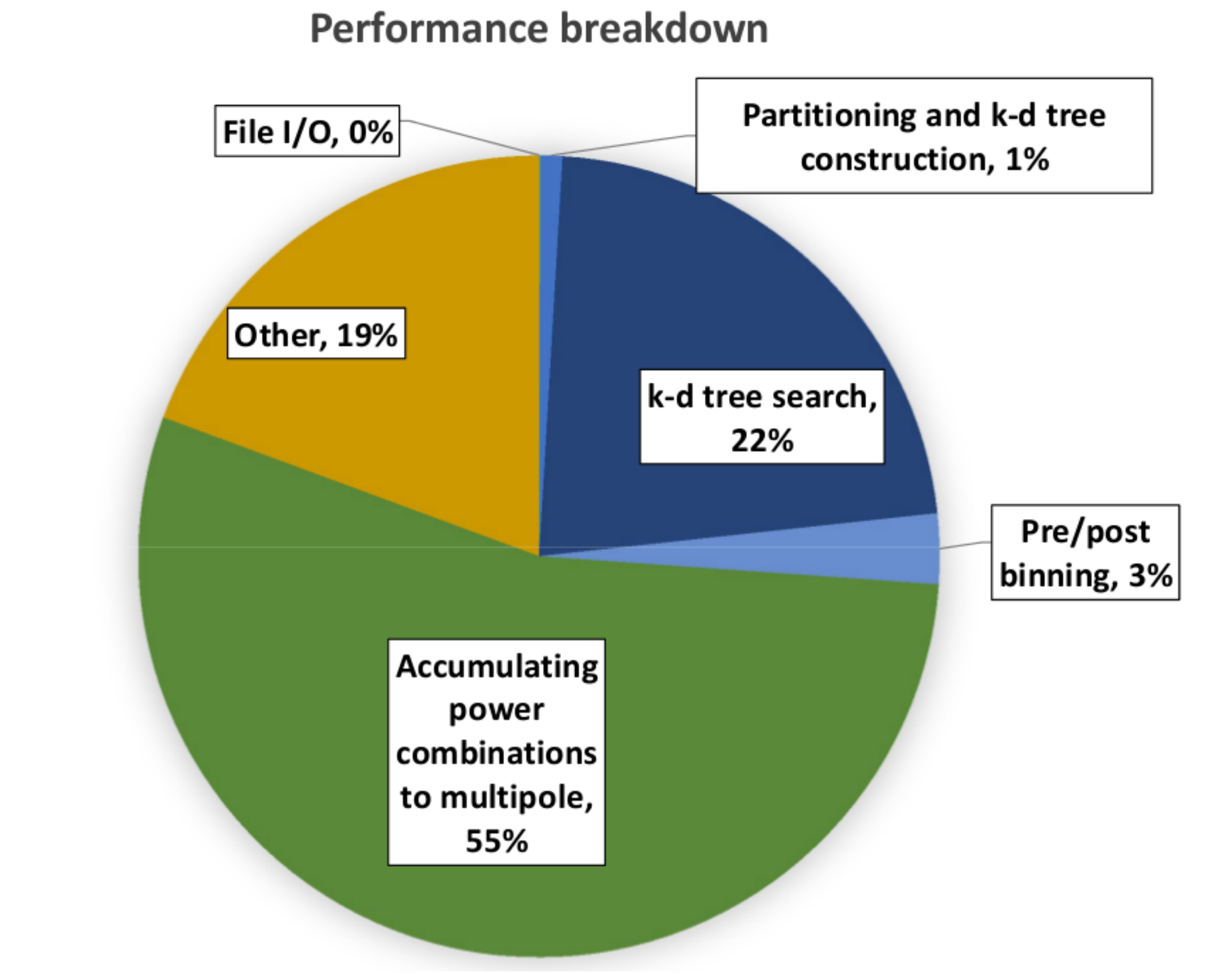}
    \caption{Runtime breakdown of single node performance on the Outer Rim 225,000 galaxy dataset.}
\label{fig:single-node}
\end{figure}

We discuss the single-node performance of Galactos here.
Figure~\ref{fig:single-node} shows the performance breakdown of the code
running the Outer Rim dataset with 225,000 galaxies on a single node. Of the
portions shown in the figure, only the k-d tree construction (which includes
partitioning and halo exchange) involves MPI communication. Hence, the code is
bound by single node performance. We see that the majority of the time ($55\%$)
is spent in the multipole accumulation function. This function (described in
Section \ref{subsec:single-node-optimizations}) calculates the 286 multipole
contributions of each pair and is vectorized over pairs. From
\ref{subsubsec:vectorization} we see that a pair of galaxies consumes 576 FLOPS; we
measure empirically that each pair in the k-d tree search contributes roughly
37 FLOPs, leading to an average of 609 FLOPs per galaxy pair for the entire
computation. The efficiency of the single-node code is bound by the instruction
mix (ratio of store to FMA instructions) and the size of the vector register
file. The multipole accumulation function achieves 1017 GF in double precision,
which is $39\%$ of a single node's peak performance. We note that the multipole
accumulation function runs entirely in double precision, but the k-d tree
search is performed in single-precision due to its insensitivity to the precision
of galaxy locations.

\subsubsection{Thread Scaling}

\begin{figure}
\centering
\includegraphics[width=\columnwidth]{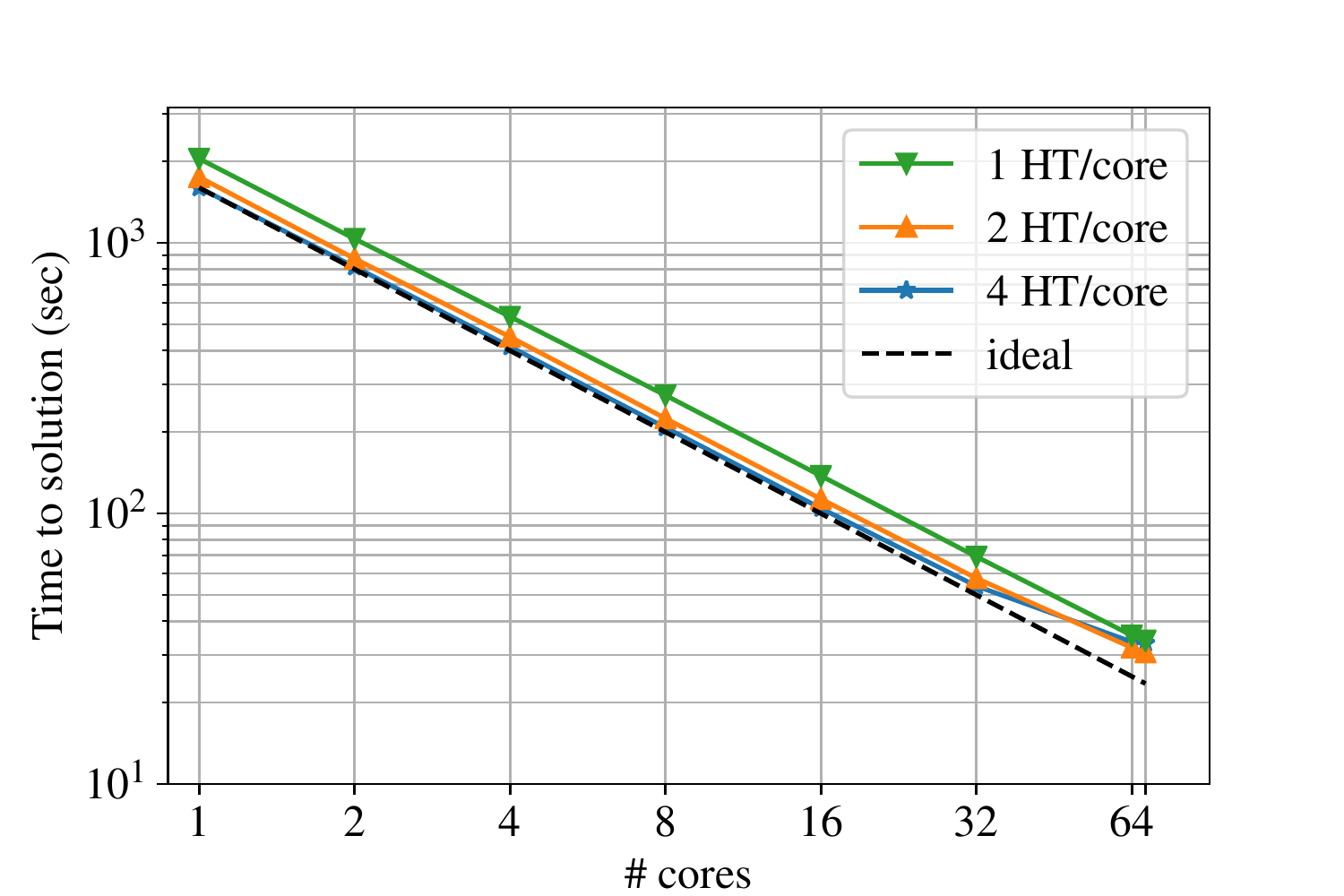}
    \caption{Thread scaling on a single Xeon Phi node with 68 cores, using 10,000 galaxies from the Outer Rim dataset. Each plot indicates a different number of hyperthreads ("HT") active on each physical core. The rightmost data point represents code performance on 68 cores, but is unlabeled on the x-axis for visual clarity.}
\label{fig:thread-scaling}
\end{figure}

Figure \ref{fig:thread-scaling} shows the thread scaling of the full application on 10,000 Outer Rim galaxies on a single node. As we increase the number of physical cores from 1 to 68, we achieve very good scaling with the time-to-solution decreasing by $58\times$. However, for a given number of physical cores, we see only a marginal improvement ($35\%$) in performance when using hyperthreading.
In fact, the performance of the k-d tree search deteriorates slightly when using hyperthreading. Overall, we achieve $65\times$ thread scaling when comparing 272 threads to 1 thread.

\subsection{Weak scaling}
\label{sec:weak-scaling}

The performance of the Galactos algorithm is sensitive to the average number density of galaxies.
Therefore, in order to capture the ``true'' performance behavior of the algorithm on smaller problem sets for weak scaling measurements, we constructed problem sets with the same number density as the full Outer Rim dataset (roughly 0.071 galaxies $[\text{Mpc}/h]^{-3}$). 
The sizes of these problem sets are shown in Table~\ref{tab:weak_scaling_problem_sets}.

\begin{figure}
\centering
\includegraphics[width=\columnwidth]{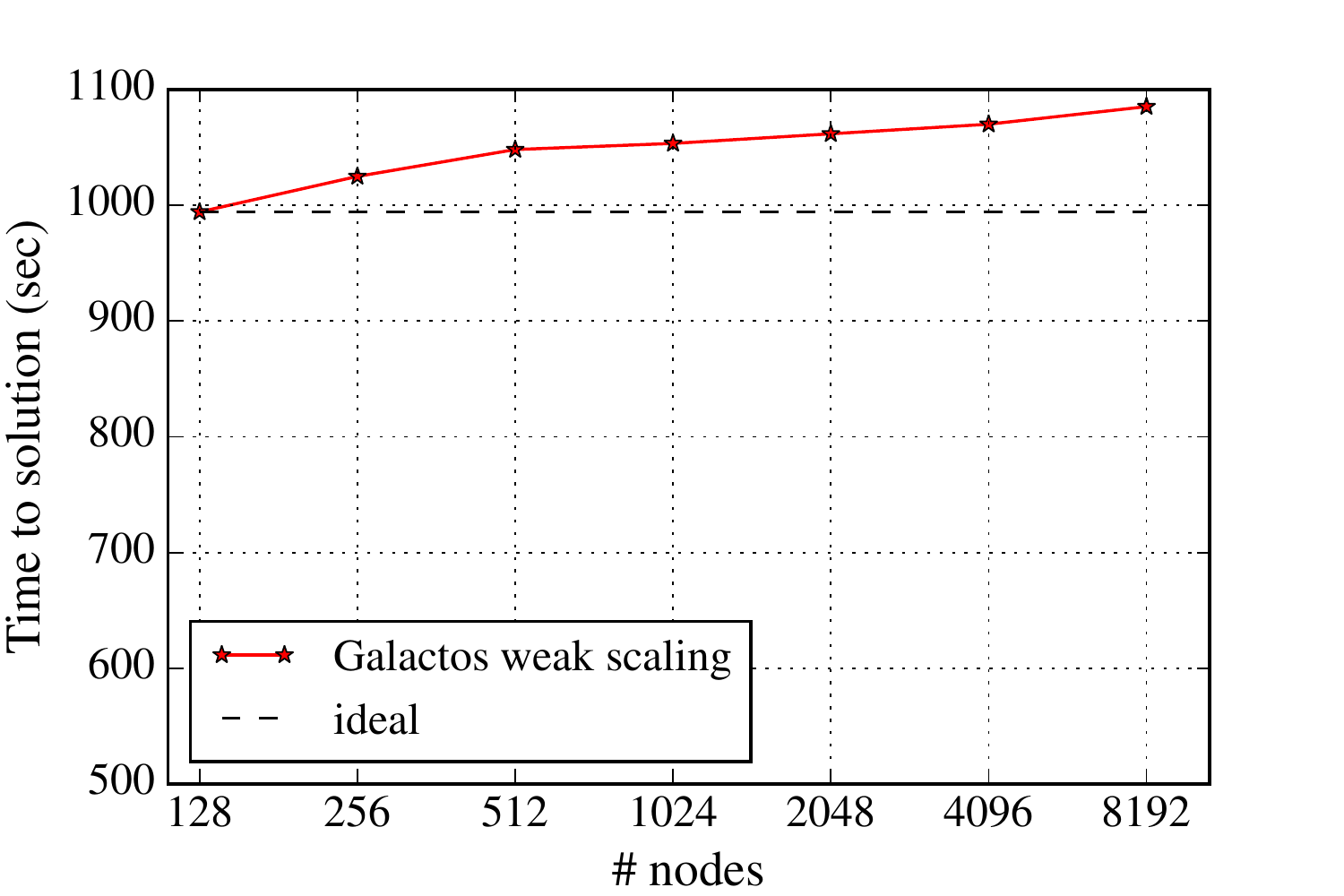}
    \caption{Weak scaling of Galactos code on Cori, using the Outer Rim datasets shown in Table~\ref{tab:weak_scaling_problem_sets}.
             The wall clock times reported here measure end-to-end code execution time, including I/O, k-d tree construction and querying, and multipole computation.}
\label{fig:weak-scaling}
\end{figure}

Figure~\ref{fig:weak-scaling} depicts the weak scaling behavior of Galactos on Cori.
With a 32-fold increase in the number of nodes---from 128 to 8,192 nodes---the time to solution increases by 9\%.
That Galactos exhibits such efficient weak scaling is expected, since is it designed to decompose the problem in order to eliminate communication between MPI processes during the bulk of the computation (as described in Section \ref{subsec:multinode_scaling}).

\subsection{Strong scaling}
\label{sec:strong-scaling}

\begin{figure}
\centering
\includegraphics[width=\columnwidth]{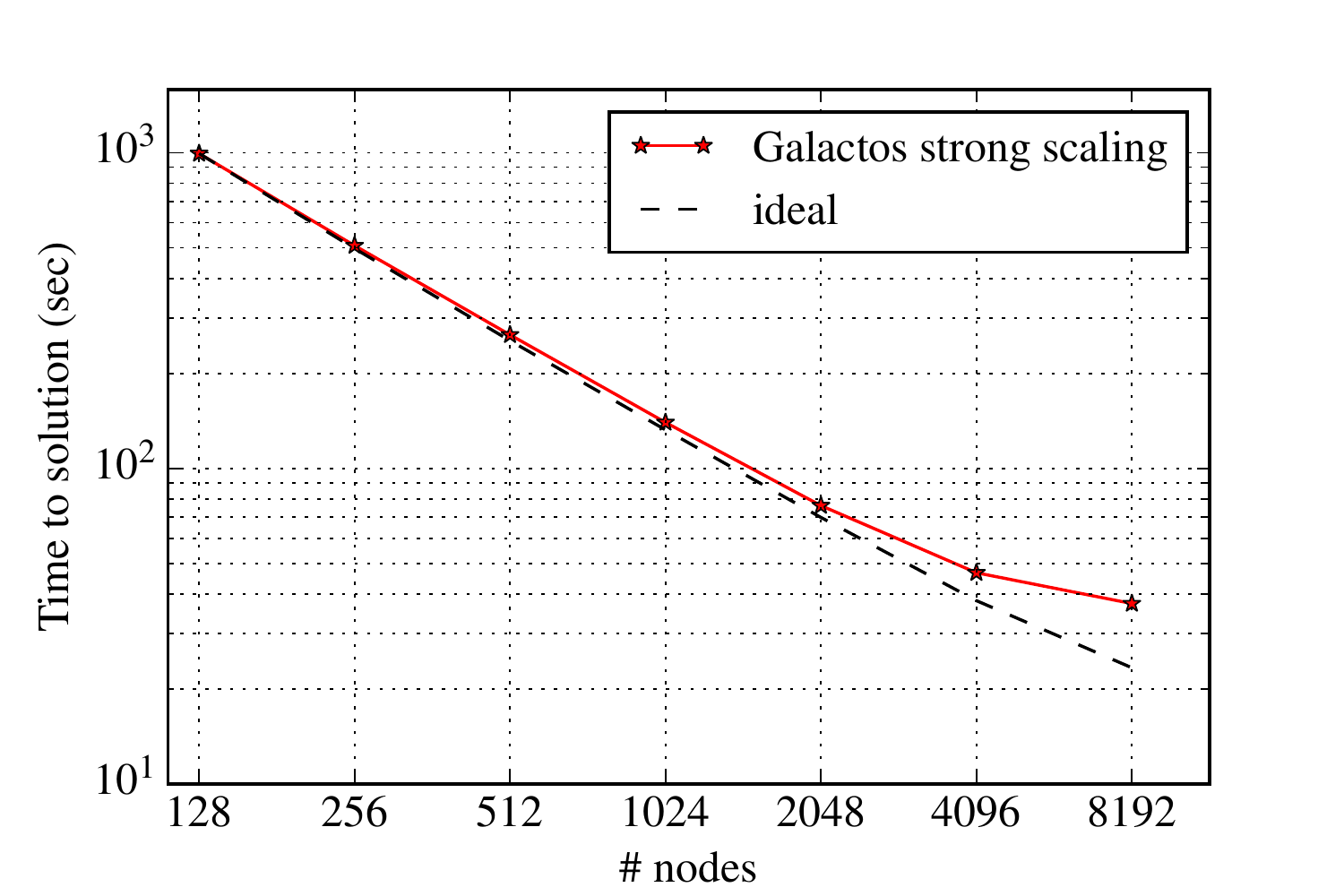}
    \caption{Strong scaling of Galactos code on Cori, using the Outer Rim dataset corresponding to 128 nodes (28.8 million galaxies; see Table~\ref{tab:weak_scaling_problem_sets}).}
\label{fig:strong-scaling}
\end{figure}

To measure the strong scaling behavior of Galactos, we use the problem set created for 128 compute nodes (28.8 million galaxies, $734.5~\text{Mpc}/h$ cubic box length; see Table~\ref{tab:weak_scaling_problem_sets}), and compute the 3PCF using 128 to 8,192 nodes.
The results are shown in Figure~\ref{fig:strong-scaling}.

We see that the algorithm strong scales efficiently.
Increasing the number of nodes by 64$\times$ (128 to 8,192 nodes) reduces the time-to-solution by a factor of 27 (994s to 37s).
This result illustrates the efficient load balancing scheme of the k-d tree implementation in Galactos. 
The small deviation from perfect strong scaling is largely due to load imbalances induced by short range variations in the galaxy density---the number of primary galaxies per node was balanced to 0.1\%, but we observed up to 60\% variation in the number of primary/secondary pairs when sub-dividing the 128-node dataset between larger numbers of nodes.
In contrast, the weak scaling runs exhibited less than 10\% variation in the number of primary/secondary pairs per node, which is consistent with the expectation that the density of galaxies becomes more uniform on larger length scales.

\subsection{Time to solution and peak performance for Outer Rim dataset}

We ran Galactos over the full Cori system on all 9,636 available nodes in both
mixed and double precision. (In mixed precision, the k-d tree is computed in
single precision and everything else is in double precision.) We can estimate
the sustained FLOPS rate for Galactos by extrapolating our empirical
measurements of FLOPS per pair from Section~\ref{sec:single-node-performance}.
The time to solution to compute the 3PCF for 2 billion galaxies in mixed
precision is 982.4 sec; in pure double precision, the time to solution is
1070.6 sec. (This is slightly faster than expected from our weak scaling study,
since we based our weak scaling datasets on a conservative estimate of $9,000$
available compute nodes. In our full-system runs each compute node consequently
processed the 3PCF for less than 225,000 primary galaxies.) In the full Outer
Rim calculation there are \num{8.17e15} galaxy pairs. Therefore, with the k-d
tree search running in double precision, we achieve an overall FLOPS rate of
4.65~PF; in mixed precision mode, we achieve 5.06~PF. There is a 9\%
improvement in overall runtime in mixed precision mode.

As a sanity check, we see that the node with the smallest number of galaxy
pairs (\num{7.06e11}) in the full run completed its multipole computation in
644.2~\si{\second}. Assuming a flop rate of 1.017~TF, we can estimate that
the node spends $61\%$ of its runtime in the multipole accumulation kernel. For
the node with the largest number of galaxy pairs (\num{9.88e11}), the
corresponding fraction is $58\%$. Note that these agree very well with the
$55\%$ measured in Section \ref{sec:single-node-performance}.

Although we obtained a precise measurement of peak performance on a single node
(1.017~TF, or 39\% of peak in double precision), extrapolating this value to
obtain overall peak application performance is non-trivial. A direct
extrapolation would suggest a peak FLOPS rate of 9.8~PF across 9636 nodes; in
reality, the actual peak was likely only marginally higher than the sustained
average of 5.06~PF (in mixed precision). This is because the multipole addition
kernel execution is highly asynchronous among nodes -- each node executes the
kernel as soon as it finds 128 galaxy pairs within its domain of the k-d tree
-- and the probability that a significant number of nodes are executing that
kernel simultaneously is vanishingly small. One could adjust the algorithm to
obtain a higher peak FLOPS rate by enlarging the galaxy ``bucket'' size such
that each node executes the multipole kernel only after it has found \emph{all}
possible galaxy pairs, thus ensuring that all nodes are executing the kernel at
once. However, this approach would also require a larger memory footprint,
introduce unnecessary synchronization, and exhibit a lower overall FLOPS rate.

\section{Discussion}

\subsection{Accelerating analysis of current and future sky surveys}

The enormous speed of Galactos offers the opportunity to substantially accelerate all analysis steps for an anisotropic 3PCF measurement. In addition to the 3PCF of the data, a full 3PCF analysis demands running the algorithm on hundreds to thousands of catalogs with spatially random clustering as well as similar numbers of mock, simulated-galaxy-data catalogs. 

Astronomical surveys of the sky have many blind spots. For example, they cannot see through the dense center of the Milky Way, or identify galaxies behind the glare of a bright star. Further, the distance to which they can observe galaxies varies over the sky due to atmospheric fluctuations and instrumental effects. The survey geometry therefore ends up very different from the perfect cube of simulated data used in this work. Consequently, in addition to computing the 3PCF of the data, one must also compute the 3PCF of hundreds of full-survey-size random catalogs, which Monte-Carlo sample the complicated survey geometry. These random catalog results enable the removal of spurious signal generated by the survey geometry. This correction step is key for extracting constraints on cosmological parameters from the data. 

Full survey-size mock catalogs, generated from simulations such as Outer Rim, are equally important for a full 3PCF analysis. First, they enable end to end verification of the analysis pipeline for systematic errors -- a simulation with known input parameters can be analyzed to check that the output parameters from fitting a 3PCF model faithfully recover the inputs. Second, the mocks can be used to test whether the theoretical models fully capture the underlying physics we believe affects the 3PCF. These models' goodness of fit to a large sample of mocks (necessary to reduce random scatter) helps determine which models should be fit to the data, and on what physical scales different models apply.

Further, a large number of mocks are often required to estimate the 3PCF's covariance matrix, which describes how independent the measurements of different triangles are from each other. This matrix needs to be inverted to optimally weight the data when fitting a model to it, and the inverse can be highly sensitive to random scatter introduced if one does not use a large number of mocks. Indeed, modern redshift surveys spend most of their compute time calculating clustering statistics on an enormous number of mocks to derive smoothly invertible covariance matrices. While other approaches to the covariance do exist \cite{SE3ptalg}, this is a standard technique that Galactos will accelerate tremendously for the 3PCF. 

An alternative estimation technique that Galactos also enables is the use of jack-knifing to estimate the small-scale covariance matrix for the 3PCF. Partitioning the survey spatially to parallelize over many nodes amounts to jack-knifing: retaining the local 3PCF results on a per node basis would therefore constitute many samples of the 3PCF over small volumes. These can be combined to provide a covariance matrix. 

\subsection{Upcoming datasets and possible science return with Galactos}

Current astronomical datasets already consist of roughly 1 million galaxies, a problem tractable in seconds on Cori with Galactos. In the next decade, these datasets will expand by orders of magnitude with surveys such as DESI, LSST, Euclid, and WFIRST. Galactos' speed will be essential for enabling the full scientific return from these datasets, and given the need for random and mock catalogs as well. 

A full description of the likely science returns of using the anisotropic 3PCF via Galactos with these enormous datasets is beyond the scope of this work, but to give a sense of what can be expected we quote some rough estimates based on our Outer Rim dataset. A survey with similar properties to Outer Rim would likely increase the precision on cosmological constraints relative to BOSS, the present state-of-the art galaxy survey, by a factor of roughly 100.  This would help cosmologists close in on the true nature of dark energy, be an unprecedentedly precise probe of a complete theory of gravity, and offer a stringent test of our best current models of galaxy formation.

\subsection{Prospects outside cosmology}
We stress that that the anisotropic 3PCF framework applies beyond just cosmological datasets. The core algorithm can be applied to any point set and can also be generalized to gridded data, enabling further acceleration. One very different astronomical application of the 3PCF is studying the properties of the Inter Stellar Medium (ISM), birthplace of stars. 
Physical conditions in the ISM such as magnetic fields, turbulence, and shocks move around dust that can be observed, and it has been shown that the 3PCF's Fourier-space analog (bispectrum) is therefore a sensitive probe of these conditions \cite{Burkhart09}. Galactos will be particularly enabling because the large parameter space of different physical conditions means one needs to compute the 3PCF of many simulations.

Any application that has translation-invariant clustering of a point set in a known but possibly irregular volume could consider these methods as a way to expand from a two-point to a three-point analysis, and many physical situations will generate non-Gaussian randomness in which the 3PCF contains new information.  Simple alterations to the algorithm  enabling use with 2-D data (e.g. generalizing \cite{SE3ptalg}) or in more than 3-D are also possible (e.g. using the higher-dimensional spherical harmonics of \cite{Frye2012}).

\section{Conclusions}
In this paper we have presented Galactos, an algorithm to calculate both the isotropic and anisotropic 3PCF for the largest simulation dataset available (of about 2 billion halos) in 20 minutes on Cori. 
Data partitioning is performed by a scalable k-d tree, resulting in good load-balancing. 
Its single-node performance has been highly optimized for Intel Xeon Phi, reaching 39\%\ of peak, with efficient use of vectorization and the full memory hierarchy. 
Galactos presents almost perfect weak- and strong-scaling, and achieves a sustained 5.06~PF across 9636 nodes.

The enormous speed of Galactos offers the opportunity to substantially accelerate all analysis steps for an anisotropic 3PCF measurement. In addition to the 3PCF of the data, a full 3PCF analysis demands running the algorithm on hundreds to thousands of catalogs with spatially random clustering. These random catalog results enable the removal of spurious signal generated by the survey geometry. 
This correction step is key for extracting constraints on cosmological parameters from the data, but increases the amount of computation required to solve the science problem by orders of magnitude compared to calculating the 3PCF for the data alone. 

Current observational datasets consist of roughly 10 million galaxies, a problem solvable in seconds on Cori with Galactos. In the next decade, these datasets will expand by several orders of magnitude to over 10 billion galaxies with surveys such as DESI, LSST, Euclid, and WFIRST. Galactos' speed (and scaling capabilities) will be essential for enabling the full scientific return from these datasets. 
In short, the algorithmic and computational developments of Galactos make the 3PCF of all future astronomical surveys for the next 20-30 years a solved problem with HPC. 

Furthermore, even with current resources, Galactos would enable computation of the 3PCF for all galaxies in the observable Universe (100 billion) in less than a day.

\section{Acknowledgments}
This research used resources of the National Energy Research Scientific Computing Center (NERSC), a DOE Office of Science User Facility supported by the Office of Science of the U.S. Department of Energy under Contract No. DE-AC02-05CH11231, and of the ALCF, which is supported by DOE/SC under contract DE-AC02-06CH11357. 
We would like to thank Doug Jacobsen, Tina Declerck, David Paul and Rebecca Hartman-Baker for assisting with Cori reservations, and Katrin Heitmann at the Argonne Leadership Computing Facility for providing the Outer Rim simulation dataset.
We thank Shirley Ho for useful conversations.
ZS acknowledges support from a Chamberlain Fellowship at Lawrence Berkeley National Laboratory and from the Berkeley Center for Cosmological Physics. DJE acknowledges support as a Simons Foundation Investigator and
from U.S. Department of Energy grant DE-SC0013718.

\bibliographystyle{ACM-Reference-Format}

\bibliography{reference}

\end{document}